%% file: pagliaroli.tex
\documentclass[vecphys]{svmult}

\usepackage{makeidx}         
\usepackage{graphicx}        
\usepackage{multicol}        
\usepackage[bottom]{footmisc}

\makeindex             

\begin{document}

\title*{Analysis of Neutrino Signals from SN1987A}
\author{G.Pagliaroli\inst{1}, M.L. Costantini\inst{1}\and
F.Vissani\inst{2}}
\institute{University of L'Aquila, Coppito (AQ), Italy \\
\texttt{giulia.pagliaroli@lngs.infn.it
\\
marialaura.costantini@lngs.infn.it} \and INFN, Laboratori Nazionali del Gran Sasso, Assergi (AQ), Italy \texttt{vissani@lngs.infn.it}}
%
%

\maketitle
\begin{abstract}
{We study SN1987A neutrino events through a likelihood analysis with one-component (\emph{cooling}) and two-component (\emph{accretion} and
\emph{cooling}) emission model. We show that there is a $3.2\sigma$ hint for the initial accretion phase.}
\end{abstract}

\section{Introduction}

On $23^{rd}$ February  $1987$ four neutrino detectors collected a burst of events from supernova (SN) explosion. The signal was detected by
Kamiokande-II (KII) in Japan (11,+5 below threshold, events) \cite{kam}; IMB in Michigan (8 events) \cite{imb} and Baksan in Russia (5 events)
\cite{baksan}, for an amount of 29 neutrino events in a window of $T=30$ sec. Four hours before 5 other events in a time window of 7 sec were
detected by the LSD experiment \cite{lsd}.
To the best of our knowledge, the first phase of the neutrino emission, revealed by LSD, cannot be explained inside the standard description of a
core-collapse SN \cite{olgarev}, whereas the second main phase of the neutrino emission can be described using the standard scenario that we adopt in
the present work.

We investigate the neutrino events detected by KII, IMB and Baksan, in order to obtain maximal information through a detailed statistical analysis.
We reconstruct the likelihood function for all events and maximize this probability varying some theoretical parameters related to the emission
models that we take into account. In our analysis, we consider the time-energy distribution of the signal, the directional information of the data,
the detectors properties (e.g. the efficiency functions). Furthermore we include a detailed description of the background, following and improving
the seminal work of Loredo and Lamb \cite{loredo,noi}.

\section{Emission models}
\label{sec:2} To grasp the emission models used in our data analysis, now we briefly describe the main phases of the so-called standard model for a
core-collapse SN \cite{janka2}. At the end of its life, a massive star consists of a sequence of concentric shells corresponding to the relics of
different burning phases and its inner core is formed by iron, which is the final stage of nuclear fusions. The iron core grows due to silicon shell
burning and, when it reaches a mass of about $1.44M_\odot$ (the Chandrasekhar limit mass), the electron degeneracy pressure can not support the
structure's weight and then the core collapses. At the densities and temperatures involved, the processes of electron capture, $\beta$ decay and
partial photodisintegration of iron-group nuclei to alpha particles occur and cause the acceleration of the collapse and the neutrino trapping in the
core. This collapse proceeds until nuclear densities of about $10^{14}g/cm^3$ are reached. At this density the nuclear matter is nearly
incompressible and the outer iron core rebounds driving a shock wave that propagates outwards, whereas the external region falls inwards at
supersonic speed. The explosion mechanism is still uncertain but, in `delayed scenario' \cite{janka}, the shock seems to loose its energy because of
the dissociation of heavy nuclei into nucleons and because of the neutrino emission, that grows when the shock crosses the neutrinosphere.
The weakened shock stagnates and transforms into a standing accretion shock, whereas the outside matter falls inward and joins the nascent compact
remnant. We call this phase \emph{accretion} (suffix {\em a}) phase and we suppose that, in this phase, $\nu_{e}$ and $\bar{\nu}_{e}$ are produced in
similar amount by the $e p\to n \nu_e$ and $e^+ n\to p \bar\nu_e$ processes.
The accretion phase occurs within the first second of $\bar\nu_e$ emission, and for this non-thermal phase \cite{nad} we assume the following
parameterized neutrino flux
\begin{equation}
\Phi^{0}_{acc}(t,E_\nu)= \frac{1}{4\pi D^2} \frac{\pi c}{(h c)^3} \left[ 8 \frac{\varepsilon(t) Y_n M_a  }{m_n} \ g(E_\nu,T_a)\ \sigma_{e^+n}(E_\nu)
\right], \label{flusso}
\end{equation}
where $Y_n=0.6$, $M_a$ is the accreting mass exposed to the positrons thermal flux, $g(E,T)=E^2/\left[1+\exp \left( E/T \right) \right]$, a
Fermi-Dirac distribution with temperature $T_a$, $\sigma_{e^+n}(E_\nu)$ is the cross section of positron interactions increasing quadratically with
$E_\nu$. The time scale of accretion process (namely $\tau_a$) appears in the following function
$\varepsilon(t)=\exp[-(t/\tau_a)^{10}]/[1+t/(0.5\mbox{ s})]$.\\ Taking into account the neutrino oscillations and the accretion assumptions (i.e.
$\Phi^{0}_{accr,\bar\nu_{\mu}}=0$), the $\bar\nu_e$ accretion flux is reduced by the factor $P_{ee}=cos^2\theta_{12}$ (the survival probability
function \cite{fogli,strumia}), hence the total flux becomes
\begin{equation}
\Phi_{acc}(t,E_\nu)=P_{ee}\cdot\Phi^{0}_{acc,\bar\nu_e}(t,E_\nu). \label{acc}
\end{equation}
%
The nascent proto-neutron star evolves in a neutron star (with radius $R_{NS}$) and this process is characterized by an intense flux of all the
species of (anti)neutrinos. We call this phase \emph{cooling} phase (suffix {\em c}), a thermal phase with a longer time scale, and we suppose that
an equal amount of energy goes in each species (equipartition hypothesis). The adopted form of the parameterized antineutrino flux, differential in
the energy is
\begin{equation}
\Phi^0_{cool}(t,E_\nu)= \frac{1}{4\pi D^2} \frac{\pi c}{(h c)^3} \left[ 4\pi R^2_c\ g(E_\nu,T(t)) \right], \label{flusso}
\end{equation}
that is a standard black body emission from the neutrinosphere with radius $R_c \simeq R_{NS}$. The time scale of cooling emission $\tau_c$ is
included in the function $T(t)=T_c\ \exp[{-t/(4\tau_c)}]$. In this phase all species of neutrino are emitted and we have to consider the neutrino
oscillations to obtain the total $\bar\nu_e$ flux at the detectors. Assuming equipartition and normal hierarchy for neutrino masses, the total
cooling flux of the electron antineutrino is
\begin{equation}
\Phi_{cool}(t,E_\nu)=P_{ee}\Phi^0_{cool, \bar{\nu}_e}(t,E_\nu)+(1-P_{ee})\Phi^0_{cool,\bar{\nu}_\mu}(t,E_\nu). \label{cool}
\end{equation}

\section{Data analysis}
\label{sec:3} Let us construct the likelihood function for the data set. We assume that the detected $\bar\nu_e$ interact for Inverse Beta Decay
(IBD), $\bar\nu_e p\to n e^{+}$. The signal rate (triply differential in time, in the positron energy $E_e$, in the cosine of the angle $\theta$
between the antineutrino and the positron directions) is given by
\begin{equation}
S(t,E_e,\cos\theta)=N_p\ \frac{d\sigma}{d\cos\theta}(E_\nu,\cos\theta)\ \eta_d(E_e)\ \xi_d(\cos\theta)\ \Phi_{\bar{\nu}_e}(t,E_\nu)\ \frac{d E_\nu}{d
E_e}, \label{seta}
\end{equation}
where $N_p$ is the number of targets (free protons) in the detectors, $\sigma$ is the IBD cross section \cite{sez}, $\eta_d$ is the --detector
dependent--average detection efficiency, $\xi_d$ is the angular bias ($\xi_d=1$ for Kamiokande-II and Baksan whereas $\xi_d(\cos\theta)=1+0.1
\cos\theta$ for IMB \cite{bratton}) and, finally, $\Phi_{\bar{\nu}_e}$ is the total flux of $\bar\nu_e$, sum of the two term $\Phi_{acc}$ and
$\Phi_{cool}$ shown respectively in Eqs. (\ref{acc}) and (\ref{cool}). The theoretical parameters, that we have to deduce by fitting the data, are
included in this last term and are 6 parameters (3 for each phase): $M_a$, $T_a$ and $\tau_a$ for accretion; $R_c$, $T_c$ and $\tau_c$ for cooling.
The likelihood function is
\begin{equation}
\label{LLL} L=\prod_{d=k,i,b}{\cal L}_d,
\end{equation}
where the suffix $d$ ranges over the detectors. Using the Poisson statistic, the likelihood function for each detector is
\begin{equation}
{\cal L}_d= e^{-f_d \int_{-t_d}^{T}\!\!  S(t+t_d) dt} \prod^{N_d}_{i=1} e^{S(t_i+t_d) \tau_d  } \left[ \frac{B_i}{2}\! +\! \int\!\! S(t_i+t_d,
E_e,c_i) G_i(E_e ) dE_e \right], \label{lik}
\end{equation}
where $N_d$ is the number of events for each detector and the suffix $i$ refers to the $i$-th event ($i=1...N_d$). The time $t_d$, called ''offset
time'', is the temporal gap between the arrival of the first neutrino to the Earth and the detection of first neutrino event in the detector. As
consequence, we add 3 new parameters $t_d$ to find out by our data analysis. The term $f_d$ is the detector live fraction ($f_d=1$ for KII and
Baksan, whereas $f_d=0.9055$ for IMB), $\tau_d$ is the detector dead time ($\tau_d=0$ for KII and IMB, whereas $\tau_d=0.035$ for IMB). Using the
well known background distribution, we calculate the probability that each event is a background signal, $B_i=B(E_i)$, and the gaussian function
$G_i$ including the energy error $\delta E_i$ arising by the energy smearing. During the cooling phase, we assume that the muon and tau antineutrinos
temperatures ($T(\bar\nu_\mu)$ and $T(\bar\nu_\tau)$ respectively) are proportional with the electron antineutrino temperature ($T(\bar\nu_e)$), i.e.
$T(\bar\nu_\tau)/T(\bar\nu_e)=T(\bar\nu_\mu)/T(\bar\nu_e)=1.2$ \cite{keil}.

At first neglecting the accretion phase, we solely consider the neutrino flux of cooling phase. We study the probability function $L$ and we find a
maximum when the model parameters reach the best-fit values shown in Table \ref{tab:1}.
\begin{table}
\centering \caption{Results for one-component (cooling) model with $2\sigma$ errors}
\label{tab:1}       
\begin{tabular}{c|c|c|c|c|c}
\hline\noalign{\smallskip}
$T_c$(MeV) & $\tau_c$(sec) & $R_c$(Km) & $t_{KII}$(sec) & $t_{IMB}$(sec) & $t_{Bak}$(sec) \\
\noalign{\smallskip}\hline\noalign{\smallskip}
$4.3^{+1.3}_{-0.9}$ & $3.7^{+2.4}_{-1.6}$ & $31^{+32}_{-16}$ & $0^{+0.9}$ & $0^{+0.4}$ & $0^{+4.5}$  \\
\noalign{\smallskip}\hline
\end{tabular}
\end{table}
We remark that the best-fit value for $R_c$ is larger than the theoretically expected one (namely $R_c \simeq R_{NS} \simeq 10Km$ \cite{janka3}). We
calculate the total energy carried by neutrinos during this phase corresponding to the gravitational binding energy of neutron star ${\cal E}_b$.
Using the equipartition hypothesis the relations ${\cal E}_b=6\cdot{\cal E}_c(\bar\nu_e)=3.39\ 10^{-4} R_c^2 T_c^4 \tau_c=3.87\cdot 10^{53}erg $
hold, where the mean values of antineutrino energy are $\langle E_{\bar\nu_e}\rangle=10MeV$ and $\langle E_{\bar\nu_x}\rangle=12MeV$, a bit lower
than expected \cite{keil,bahcall}.

Motivated by the experimental fact that about $40\%$ of the SN1987A events have been recorded in the first second, we consider the accretion phase
completing the emission model. We set $M_a=0.5 M_\odot$ that is a reasonable value of the outer core mass, therefore we maximize the likelihood as a
function of the other parameters.
%
%
%
%
\begin{table}
\centering \caption{Results for two-components (accretion and cooling) model with $2\sigma$ errors}
\label{tab:2}       
\begin{tabular}{c|c|c|c|c|c|c|c}
\hline\noalign{\smallskip}
$T_c$(MeV) & $\tau_c$(sec) & $R_c$(Km) & $T_a$(MeV) & $\tau_a$(sec) & $t_{KII}$(sec) & $t_{IMB}$(sec) & $t_{Bak}$(sec) \\
\noalign{\smallskip}\hline\noalign{\smallskip}
$5.1^{+2.1}_{-1.4}$ & $4.4^{+3.6}_{-1.9}$ & $13^{+18}_{-8}$ & $2.1^{+0.2}_{-1.4}$ & $0.7^{+1.3}_{-0.3}$ & $0^{+0.8}$ & $0^{+0.7}$ & $0^{+0.6}$  \\
\noalign{\smallskip}\hline
\end{tabular}
\end{table}
\begin{figure}
\begin{center}
\includegraphics[width=0.45\linewidth]{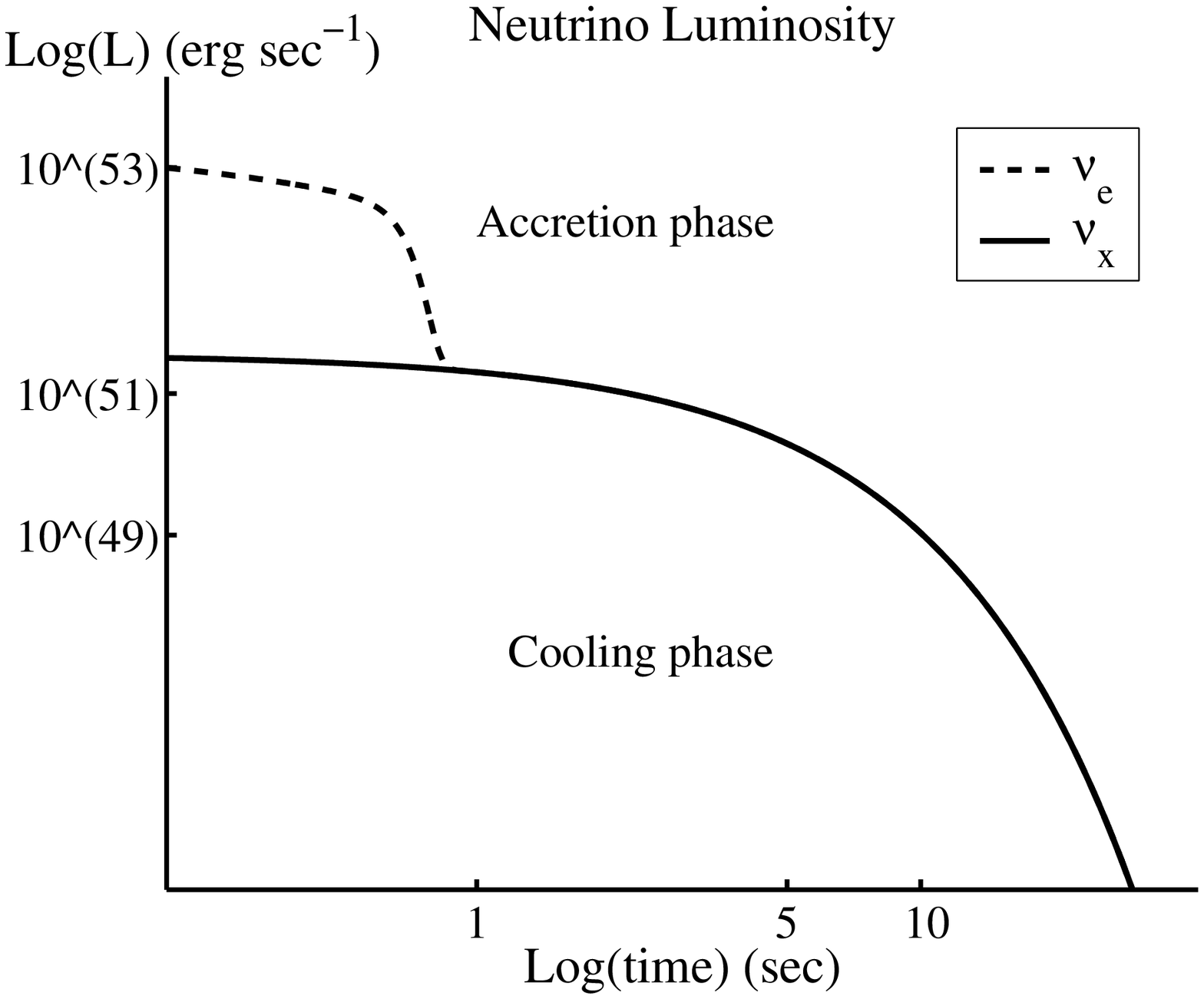}
\includegraphics[width=0.45\linewidth]{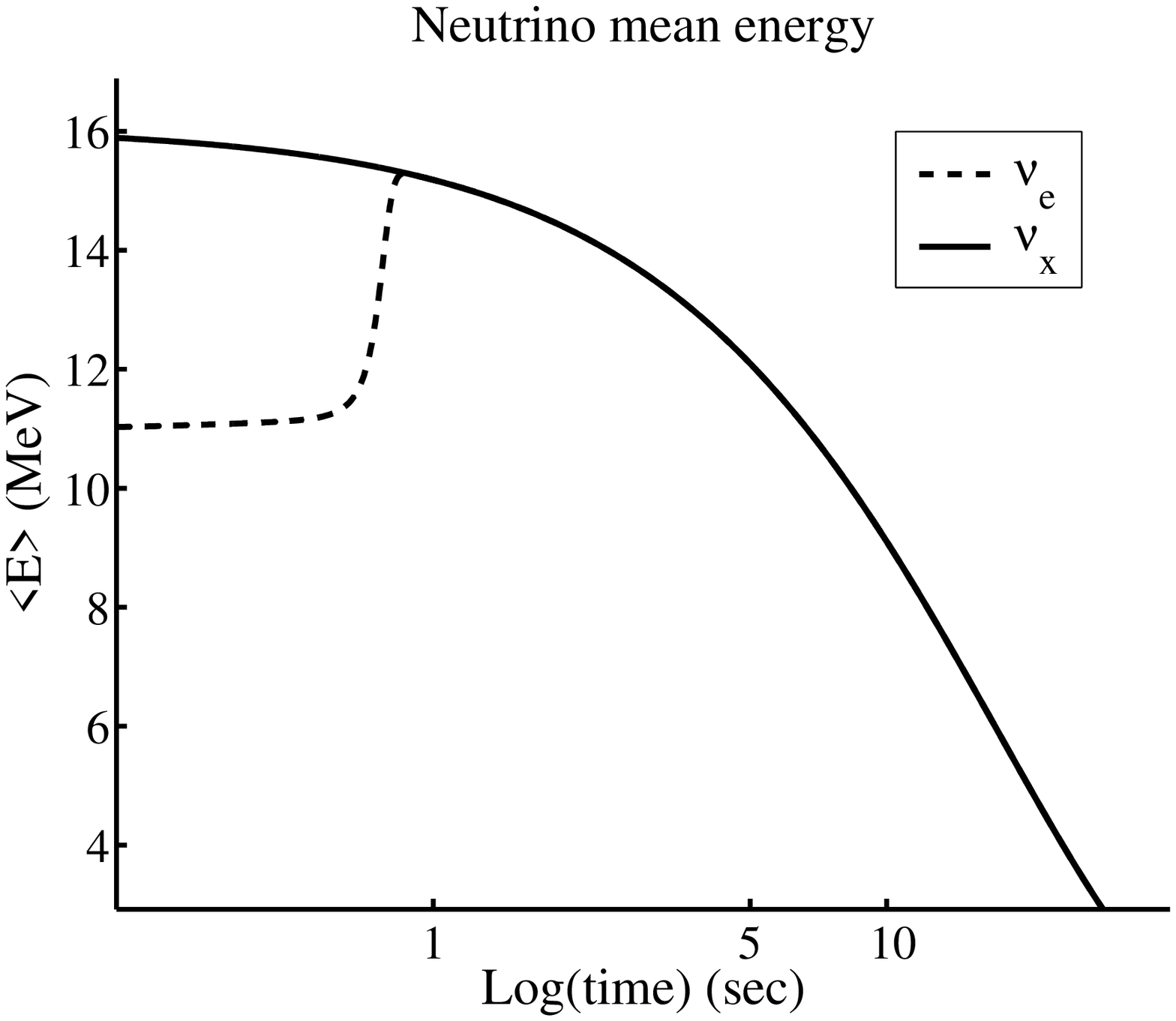}
\end{center}
\caption{\em The luminosity (a) and the mean energy (b) of $\bar\nu_e$ and $\bar\nu_x$ (dashed line and solid line, respectively) obtained from the
emission model (accretion and cooling) exploiting the best fit values of Table \ref{tab:2}.}
\label{fig:1}       
\end{figure}
\begin{figure}
\centering
$$\includegraphics[width=0.38\textwidth, height=0.41\textwidth]{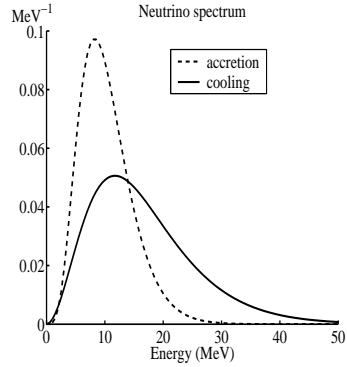}$$ \vskip-4mm
\caption{\em Time integrated energy spectra of neutrino for accretion (dashed line) and cooling (solid line) emission phase. }
\label{fig:2}       
\end{figure}
We find the best-fit values shown in Table \ref{tab:2}. Note that $T_c$, $R_c$ and $\tau_a$ are very close to theoretical values expected
\cite{janka3}. The binding energy is the sum of two terms, the energy of neutrino emitted in the cooling phase, ${\cal E}_c=1.76\cdot 10^{53}erg$,
and the energy ${\cal E}_a=2\cdot{\cal E}_a(\bar\nu_e)=4.14 M_a T_a^6 \tau_a \varphi=6.3\cdot 10^{52}erg$ carried by $\nu_e$ and $\bar\nu_e$ in the
accretion phase, where $\varphi\equiv\int_0^\infty dx\ \exp(-x^{10})\! /(1+x\ \tau_a/0.5)$. Hence we obtain ${\cal E}_b=2.4\cdot 10^{53}erg$,
$\langle E_{\bar\nu_e}\rangle_a=10.3MeV$, $\langle E_{\bar\nu_e}\rangle_c=12.6MeV$ and $\langle E_{\bar\nu_x}\rangle=15MeV$. We report the neutrino
luminosities, the neutrino energies mean values (as a function of time) in Fig. \ref{fig:1} (a) and (b), respectively. Moreover, we plot the neutrino
energy spectra in Fig.\ref{fig:2}.

\section{The evidence for the phase of accretion}
\label{sec:4}  In the analysis with accretion and cooling phases, we find that the absolute value of the likelihood function (in the best-fit point)
is about 1000 times larger than the corresponding value in case of no accretion phase, giving a significant hint for an accretion phase. In fact, let
us assume as null hypothesis $H_0$ the case where the accretion is absent and compare it with the alternative hypothesis $H_1$ with an accretion
phase described by two additional parameters $T_a$ and $\tau_a$. When we add $\nu=2$ degrees of freedom, we expect that the $\chi^2$ will decrease by
a certain amount $\Delta\chi^2$. In order to determine the rejection interval for the hypothesis $H_0$, we perform a likelihood ratio test; then the
required probability distribution function is the regularized gamma function $Q(\nu/2,\Delta\chi^2/2)$. 
When we go from $M_a=0$ (no accretion) to $M_a=0.5\ M_\odot$ (our reference point) the $\chi^2$ diminishes by $\Delta \chi^2=13.4$. As consequence,
we reject the null hypothesis in favor of the hypothesis that accretion occurred with a significance of $\alpha=\exp(-\Delta\chi^2/2)=1.2\times
10^{-3}$. In Gaussian language, this amounts to $3.2\sigma$.

%
%
%
%
\input{referenc}



\end{document}

%% file: referenc.tex
%
%

%
%

%% file: pagliaroli.bbl
\begin{thebibliography}{99.}
%
%
%


\bibitem{kam}
K.~Hirata~{\it et~al}~[Kamiokande-II Collaborati\-on]: ``Observation of a Neutrino Burst from the Supernova SN1987A,'' Phys.\ Rev.\ Lett.\ {\bf 58}
(1987) 1490.

\bibitem{imb}
R. M.~Bionta {\it et al.}:
 ``Observation of a Neutrino Burst in Coincidence with Supernova SN1987A in
the Large Magellanic Cloud,'' Phys.\ Rev.\ Lett.\ {\bf 58} (1987) 1494.

\bibitem{baksan}
E.N.~Alekseev, L.N.~Alekseeva, I.V.~Krivo\-sheina and V.I.~Volchenko: Phys.Lett.B {\bf 205} (1988) 209.

\bibitem{lsd}
V.~L.~Dadykin {\it et~al}: JEPT Lett. {\bf 45} (1987) 593.
%

\bibitem{olgarev}
V.~L.~Dadykin, G.~T.~Zatsepin and O.~G.~Ryazhskaya: Sov.\ Phys.\ Usp.\  {\bf 32} (1989) 459;
O.~G.~Ryazhskaya: Phys.\ Usp.\  {\bf 49} (2006) 1017.
%
\bibitem{loredo} T. J. Loredo, D. Q. Lamb: P. R. D.
\textbf{65}, 063002 (2002)
%
\bibitem{noi} G. Pagliaroli, M. L. Costantini, A. Ianni and F. Vissani: astro-ph/ 0705.4032 (2007)
%
\bibitem{janka2} H. Th. Janka {\it et~al}: ''Theory of Core-Collapse Supernovae'', Phys. Report \textbf{442}
(2007) 38.
%
\bibitem{janka} H. Th. Janka: A \& A
\textbf{368}, 527 (2001)
%
\bibitem{nad}
D. K.~Nadyozhin, ``The neutrino radiation for a hot neutron star formation and the envelope outburst problem,'' Astrophys.\ Space Sci.\  {\bf 53}
(1978) 131.
%
\bibitem{fogli}
G. L.~Fogli, E. Lisi, D.Montanino and A. Palazzo: ``Supernova neutrino oscillations: a simple analytical approach'', Phys. Rev. D {\bf 65} (2002)
073008
%
\bibitem{strumia}
A.~Strumia and F. Vissani: ``Neutrino masses and mixings and ...'', hep-ph/0606054, a review regularly updated on the web
%
\bibitem{sez}
A.~Strumia and F. Vissani: ``Precise quasielastic neutrino nucleon cross section'', Phys. Lett. B {\bf 564} (2003) 42
%
\bibitem{bratton}
C. B.~Bratton {\it et~al}: ``Angular distribution of events from SN1987A'', Phys. Rev. D {\bf 37} (1988) 3361
%
\bibitem{keil}
M. T.~Keil, G. G. Raffelt and H. T. Janka: ``Monte Carlo study of supernova neutrino spectra formation'', Astrophys. J. {\bf 590} (2003) 971
%
\bibitem{janka3}
H. T.~Janka {\it et~al}: ``Neutrinos from type II supernovae and the neutrino driven supernova mechanism'', Vulcano 1992 Proceedings, 345-374
%
\bibitem{bahcall}
J. N.~Bahcall: cap 15 of ``Neutrino astrophysics'', Cambridge University Press, 1989
%

\end{thebibliography}
